\documentclass[aps,prd,twocolumn,showpacs,superscriptaddress,preprintnumbers, floatfix,nofootinbib]{revtex4-2}
\usepackage{amsmath,amssymb}
\usepackage{epsfig}
\usepackage{hyperref}
\usepackage{color}
\usepackage{slashed}
\usepackage{placeins}

\usepackage{amsfonts}
\usepackage{graphicx, rotating}
\usepackage{epstopdf}
\usepackage{latexsym}
\usepackage{rotating}
\usepackage{braket}

\usepackage[export]{adjustbox}
\usepackage[font={small}]{caption}   

\usepackage[starfontserif]{starfont}

\newcommand{\Ecal}{{\mathcal E}}

 \newcommand{\Lcal}{{\mathcal L}} 
 
\newcommand{\Ocal}{{\mathcal O}}

\definecolor{myred}{rgb}{0.7, 0, 0}
\definecolor{myblue}{rgb}{0, 0, 0.7}
\definecolor{mygreen}{rgb}{0.04, 0.7, 0.5}
\definecolor{mygray}{rgb}{0.1, 0.1, 0.1}

\definecolor{oleg}{RGB}{0, 153, 76}

\hypersetup{colorlinks,citecolor=myred,linkcolor=myblue,urlcolor=myblue,linktocpage=true}

\def\be   {\begin{equation}}   \def\ee   {\end{equation}}
\def\ba   {\begin{array}}      \def\ea   {\end{array}}
\def\bea  {\begin{eqnarray}}   \def\eea  {\end{eqnarray}}
\def\bean {\begin{eqnarray*}}  \def\eean {\end{eqnarray*}}

\def\bry{\begin{array}}
	\def\ery{\end{array}}

\def\Mpl{ M_{\rm Pl}}

\newcommand{\skipnew}[1]{}

\def\SU{\small{The Oskar Klein Centre, Department of Physics, Stockholm University, 10691 Stockholm, Sweden}}

\def\Hamburg{\small{II. Institute for theoretical Physics, Hamburg University, Luruper Chaussee 149, 22761 Hamburg, Germany}}

\def\DESY{\small{Deutsches Elektronen-Synchrotron DESY, Notkestr. 85, 22607 Hamburg, Germany}}

\begin{document}

\date{\today}
\title{\Large\bfseries Detectability of accretion-induced bosenovae in the Milky Way}

%


\author{Dennis Maseizik}
\email{dennis.maseizik@desy.de}
\affiliation{\Hamburg}

\author{Joshua Eby}
\email{joshaeby@gmail.com}
\affiliation{\SU}

\author{Hyeonseok Seong}
\email{hyeonseok.seong@desy.de}
\affiliation{\DESY}

\author{Günter Sigl}
\email{guenter.sigl@desy.de}
\affiliation{\Hamburg}

\begin{abstract}
We estimate collapse rates of axion stars in our galaxy based on the axion minicluster mass function of the Milky Way dark matter halo.
We consider axion-like particles with different temperature evolution of the axion mass, including the QCD axion with $m_a=50\,\mu$eV.
Combining estimates for the present-day axion star mass function from our previous work with the axion star accretion model predicted by self-similar growth, we can infer the expected number of bosenovae occurring within the Milky Way.
Our estimates suggest that for an observation time of $t_\mathrm{obs}=1\,$yr, the majority of the up to $\sim 10^{13}$ bosenovae per galaxy occur in the densest miniclusters with initial overdensity parameter $\Phi\lesssim 10^4$.
We discuss the detectability of such recurring axion bursts within our galactic vicinity and find that, for models with derivative couplings including axion-fermion interactions, potential broadband axion DM experiments can probe a large range of ALP masses $m_a\lesssim 10^{-6}\,$eV and with moderate improvements even the QCD axion case.
For axions with non-derivative-type interactions like the axion-photon coupling, our analysis suggests that optimistic predictions with order-one dark matter abundance of axion stars $f_\star \sim 1$ can be probed by dedicated burst searches.
\end{abstract}

\preprint{DESY-24-151}

\maketitle

\section{Introduction}

The existence of beyond-Standard-Model (BSM) fields with masses below the electron-volt scale represents a compelling solution to explain the identity of dark matter (DM)~\cite{Press:1989id,Sin:1992bg,Hu:2000ke,Peebles:2000yy,Amendola:2005ad,Hui:2016ltb,Ferreira:2020fam}. 
Within this class of models, parity-odd scalar fields known as \emph{axions} and \emph{axion-like particles} (ALPs) are among the most well-motivated candidates~\cite{Peccei:1977ur,Weinberg:1977ma,Wilczek:1977pj,Kim:1979if,Shifman:1979if,Zhitnitsky:1980tq,Dine:1981rt,Preskill:1982cy,Abbott:1982af,Dine:1982ah}.
A large number of experiments are now underway which hope to discover such fields on Earth, including those using magnetic cavities or magnetic field induced nuclear spin precession (see~\cite{Adams:2022pbo} for a recent review). Parity-even fields with small masses are also widely sought using quantum-sensing experiments (see~\cite{Antypas:2022asj}).
In this work, we classify different axion-like models based on the present-day axion mass $m_a$ and its cosmological temperature evolution. 
We consider axion-like particles, including the special case of the QCD axion-like case with $m_a \simeq 50\,\mu{\rm eV}$ and similar temperature evolution of $m_a(T)$~\cite{GrillidiCortona:2015jxo}.
In the following, we will be referring to \emph{axions} as the general class of particles including both the QCD axion and ALPs.

The dark matter in the vicinity of Earth is generally non-relativistic, with typical speed $v_{\rm dm}\simeq 10^{-3}$~\cite{Schoenrich:2009bx,Eilers_2019}, and is constrained to have an average energy density of approximately $\rho_{\rm dm}\simeq 0.4\,{\rm GeV/cm}^3$~\cite{Weber_2010,Nesti:2012zp,Bovy:2012tw,Read:2014qva}. The parameters $v_{\rm dm}$ and $\rho_{\rm dm}$ play a key role in determining the sensitivity of a given experiment to the presence of DM~\cite{Adams:2022pbo,Antypas:2022asj}. 
However, ultralight fields like axions and ALPs generically form large-scale overdensities through gravitational collapse and relaxation, which can modify this simple picture. The corresponding overdensities imply novel methods to search for, and possibly discover, light scalar fields.
Importantly, these novel methods are complementary to traditional DM search strategies.

In a well-motivated scenario, early cosmological overdensities collapse at or before matter-radiation equality, forming \emph{axion miniclusters}~\cite{Hogan:1988abc,Kolb:1993zz}. The cores of miniclusters (MCs) host further overdensities known as \emph{axion stars} (ASs)~\cite{Kaup:1968zz,Ruffini:1969qy,Colpi:1986ye} through relaxation of the field~\cite{Levkov:2018kau,Chen:2020cef,Kirkpatrick:2020fwd,Chen:2021oot,Kirkpatrick:2021wwz}. 
In this work, we study the signals arising from the gravitational collapse of axion stars, wherein they generally convert a large fraction of their total mass-energy from non-relativistic axions to relativistic ones~\cite{Eby:2016cnq,Levkov:2016rkk}, a process known as a \emph{bosenova}. As a result, the energy density observed by an experimental search can be enhanced significantly compared to the local density $\rho_{\rm dm}$, and can be distinguished from the latter by its relativistic speed $v\simeq 1$. A nearby bosenova (e.g. occurring within the Milky Way) therefore represents a viable transient target for terrestrial experiments, and the potential signal strength of a transient bosenova search has been previous explored for a variety of experiment types~\cite{Eby:2021ece,Arakawa:2023gyq,Arakawa:2024lqr}. These studies focused only on the signal, leaving the rate of bosenovae in the Milky Way an open question.

Recently, the distribution of axion star masses (which we call the axion star mass function, ASMF) was derived using Press-Schechter theory predictions for the minicluster mass function (MCMF) and the core-halo relation to set the mass of the axion star core \cite{Maseizik:2024qly}.
\footnote{Note that a significant population of axion stars may also form directly at matter-radiation equality~\cite{Gorghetto:2024vnp}.
See also \cite{Chang:2024fol} for a derivation of the axion star mass distribution using the output of cosmological simulations instead of the core-halo relation.} 
Crucially, the authors of~\cite{Maseizik:2024qly} found that a significant number of axion stars could be at or near their \emph{critical mass}, i.e. the mass at which these objects become unstable to collapse due to self-interactions. 
Each year, some fraction of these near-critical axion stars can be pushed to the critical point by mass accretion of diffuse axions from their host minicluster background~\cite{Levkov:2018kau,Eggemeier:2019jsu,Chen:2020cef,Chan:2022bkz,Dmitriev:2023ipv}.
A similar study involving different accretion scenarios, which drive the AS cores of galactic miniclusters to undergo parametric resonance was recently published in \cite{maseizik_seong_2024}.
Repeated Bosenovae have also been used to constrain axion DM models in a cosmological context using decaying dark matter constraints from CMB measurements and baryon acoustic oscillation data in Refs.~\cite{Fox_2023, Nygaard_2021}.
Our study of the galactic detectability of axion bursts is complementary to Ref.~\cite{Fox_2023}, but it also incorporates extensions through consideration of the full MC mass range, and the self-similar growth model of ASs~\cite{Dmitriev:2023ipv}.

We build upon these previous works, namely \cite{Maseizik:2024qly} for the ASMF of the Milky Way and \cite{maseizik_seong_2024} for the mass growth of galactic AS cores, motivated by the semi-analytical study in \cite{Dmitriev:2023ipv} and their self-similar attractor model.
\footnote{
Note that in this work, we are restricting our analysis to the self-similar accretion model from \cite{Dmitriev:2023ipv} and the ASMF from \cite{Maseizik:2024qly}, ignoring the other accretion scenarios suggested in \cite{maseizik_seong_2024}.
We choose this approach for simplicity and because we believe the numerical simulations of \cite{Dmitriev:2023ipv} to provide the strongest evidence yet. 
Nevertheless our work can easily be updated by improved accretion modeling and ASMF determination.
}
Using predictions from self-similar accretion and the linear growth ASMF, we provide, for the first time in the literature, a precise target for experimental searches which combines information about both the signal strength~\cite{Eby:2021ece,Arakawa:2023gyq,Arakawa:2024lqr} and the number~\cite{Maseizik:2024qly} of bosenovae which are detectable as transient events in terrestrial detectors.
To do so, we estimate the rate at which these near-critical axion stars would, through accretion from the surrounding minicluster, reach their critical mass, and thereby determine the number of bosenovae occurring in the Milky Way per unit time. By populating these bosenovae using some estimate of their number distribution, we determine how many bosenovae would occur close enough to the Earth to be detected. We will study the resulting parameter space to motivate transient searches using existing experiments, as well as new searches to probe the most promising parameters.

\section{Axions, axion stars, and bosenovae}
\label{sec:axionstars}

Axions generically arise as the pseudo-Nambu-Goldstone boson $a$ of a $U(1)$ symmetry which is spontaneously broken at a high scale $f_a$ (known as the \emph{axion decay constant})~\cite{Peccei:1977ur}. If this symmetry breaking occurs after inflation (the so-called \emph{post-inflationary scenario} for axions), $\Ocal(1)$ density fluctuations arise on scales of order Hubble radius at the cosmological temperature $T$ when the axion field starts to oscillate, $m_a(T)\gtrsim 3H(T)$. These fluctuations decouple from the Hubble flow at a redshift $z_{\rm dec}(\delta) \simeq \delta z_{\rm eq}$ where $z_{\rm eq}\simeq 3400$ is the redshift of matter-radiation equality and $\delta \equiv \rho/\bar{\rho}$ is the overdensity of a given patch relative to the cosmological average density $\bar{\rho}$. The resulting overdensities, now gravitationally bound and decoupled from Hubble flow, are known as \emph{axion miniclusters}.\footnote{Axion miniclusters can be also produced abundantly in the pre-inflationary scenario through non-standard misalignment mechanisms \cite{Arvanitaki:2019rax,Eroncel:2022efc,Chatrchyan:2023cmz}.}

The distribution of axion miniclusters has been widely studied through cosmological simulations~\cite{Kolb:1993zz,Visinelli:2018wza,Xiao:2021nkb,Ellis:2022grh,Pierobon:2023ozb,Eggemeier:2024fzs}. Of particular importance is the temperature dependence of the axion mass, which can be parametrized in the form
\begin{equation} \label{eq:m_T}
    m_a(T>\Lambda_0) \simeq m_{a,0}\left(\frac{\Lambda_0}{T}\right)^n\,,
\end{equation}
where $T$ is the cosmological temperature, $\Lambda_0$ is an energy scale set to $\Lambda_0 = \sqrt{m_{a,0}f_a}$ (analogous to the topological susceptibility of the QCD axion), and $n$ is an index determining the temperature evolution. Note that here and in what follows, we use the shorthand notation $m_a = m_{a,0}\equiv m_a(T_0)$ for the present-day axion temperature, with $T_0=2.7\,$K. The index $n=0$ indicates a temperature-independent mass, whereas for the QCD axion we take $n\simeq 3.34$ based on the interacting instanton liquid model from \cite{wantz_axion_2010},
\footnote{There are uncertainties depending on the calculation method. For example, lattice calculations yield $n\simeq 3 - 4$ \cite{Berkowitz:2015aua,Borsanyi:2016ksw}, and the dilute instanton gas model gives $n=4$ \cite{Marsh:2015xka}.}.

In the present-day, axions generally exhibit an approximately shift-symmetric potential which can be expressed as
\begin{equation} \label{eq:axionpotential}
    V(a) = m_a^2 f_a^2 \left[1 - \cos\left(\frac{a}{f_a}\right)\right]\,.
\end{equation}
The leading term gives the mass $m_a$ of the axion, and the next-to-leading term gives the dominant self-interaction term, an attractive $a^4$ interaction with dimensionless coupling $\lambda \equiv -m_a^2/f_a^2$.

The value of the temperature index $n$ has an important impact on minicluster properties. 
As we will see in the next section, the structures, which are seeded by initial fluctuations of the axion field in the early universe, become heavier for larger $n$.
Once the axion mass behavior is fixed, we can fix the value of the axion decay constant $f_a$ by requiring that the coherent oscillations of the axion provide sufficient energy density to account for the total relic density of DM, i.e. $\Omega_a h^2 \simeq 0.12$.
The total relic density in the post-inflationary scenario is contributed by the misalignment mechanism and the decay of topological defects, as parametrized in \cite{Fairbairn_2018}:
\begin{equation}\label{eq:Omega_a}
    \Omega_a = \frac{1+\beta_{\rm dec}}{6H_0^2 \Mpl^2} \frac{c_n \pi^2}{3} m_a(T_0) m_a(T_{\rm osc}) f_a^2 \left[\frac{a(T_{\rm osc})}{a(T_0)}\right]^3 \,,
\end{equation}
where $\beta_{\rm dec}$ is the ratio of the relic density contribution from the decay of topological defects to that from the misalignment mechanism, defined as $\beta_{\rm dec} \equiv \Omega_{a,{\rm dec}}/\Omega_{a,{\rm mis}}$. 
We use $\beta_{\rm dec} \simeq 2.48$ based on simulations \cite{Kawasaki:2014sqa} (see also~\cite{Gorghetto:2018myk,Gorghetto:2020qws,Buschmann:2021sdq,Kim:2024wku}), and $c_n = 2.3,2.2,2.1$ for $n=1,2,3.34$ to parametrize the effects of anharmonicities in the axion potential. The oscillation temperature $T_{\rm osc}$ is defined by $m_a(T_{\rm osc}) = 3H(T_{\rm osc})$ when the axion mass becomes relevant, $H_0=H(T_0)$ is the Hubble parameter, and $\Mpl=2.4\times 10^{18}\,{\rm GeV}$ the reduced Planck mass. 

A phenomenologically important feature of axion miniclusters is that they can host a dense core of axions in the ground-state configuration, which form a smaller self-gravitating object called an \emph{axion star}~\cite{Kaup:1968zz,Ruffini:1969qy,Colpi:1986ye,Chavanis:2011zi,Chavanis:2011zm}. Axion stars (ASs) can form from relaxation of diffuse axions to the ground state of the field on astrophysical timescales~\cite{Levkov:2018kau,Chen:2020cef,Kirkpatrick:2020fwd,Chen:2021oot,Kirkpatrick:2021wwz,Dmitriev:2023ipv,Jain:2023tsr}; however, axion stars are thought to form even more rapidly within miniclusters, within a few free-fall times, by violent relaxation of the field as miniclusters collapse (see e.g.~\cite{Eggemeier:2019jsu}). 
Numerical simulations of isolated miniclusters suggest that each MC can host up to a single axion star, whose mass $M_\star$ satisfies a \emph{core-halo relation} of the form~\cite{Schive:2014dra,Schive:2014hza,Eggemeier:2019jsu,Eggemeier:2019khm}
$M_\star \propto \mathcal{M}^{1/3}$
where $\mathcal{M}$ is the mass of the total AS-MC system (see Sec. \ref{sec:accretion} for details).

In general, axion stars are stabilized through balance between the gradient pressure, the self-gravity of the axion field and the axion self interactions, which we assume to be weak and attractive throughout this work $(\lambda <0)$~\cite{Chavanis:2011zi,Chavanis:2011zm,Eby:2014fya}. Once the mass grows to a critical value, attractive self-interactions in the potential of Eq.~\eqref{eq:axionpotential} destabilize the star, leading to gravitational collapse~\cite{Chavanis:2016dab,Helfer:2016ljl}. This occurs when $M_\star$ and the corresponding radius $R_\star$ approach
\begin{align}
M_{\star,\lambda } &= \sqrt{\frac{3 }{G}} \frac{2 \pi  f_a}{m_a }  \quad \,, \quad
R_{\star,\lambda} = \sqrt{\frac{3}{32 \pi G}} \frac{1}{m_a f_a} \label{eq:M_Star_Max_R_Star_min} \,,
\end{align}
where $G$ is the gravitational constant, $M_{\star,\lambda}\propto 1/\sqrt{|\lambda|}$, and $R_{\star,\lambda}\propto \sqrt{|\lambda|}/m_a^2$. During the collapse, as the axion star becomes more dense, number-changing interactions in the core of the star rapidly convert a fraction of the non-relativistic axions into relativistic ones, which escape the star\footnote{If the axion-photon coupling is large enough, the dominant conversion process is to radio photons through parametric resonance rather than relativistic axions, as shown in~\cite{Levkov:2020txo}; we will not consider such a case for the purposes of this work.}, a process called a \emph{bosenova}~\cite{Eby:2016cnq,Levkov:2016rkk}. 
We summarize the signals produced from bosenovae in Sec. \ref{sec:signals} after introducing the corresponding accretion rates, which drive the AS cores to reach the point of criticality in Sec. \ref{sec:accretion}.

\section{Axion Star Accretion Rates and Bosenova Numbers}
\label{sec:accretion}

In the post-inflationary scenario, axion miniclusters serve as a natural starting point for the formation and evolution of axion stars. 
Drawing on linear growth predictions for the MC mass distribution, the core-halo relation by \cite{Schive:2014dra}, and the general properties of axion stars, Ref.~\cite{Maseizik:2024qly} determined the present-day distribution of miniclusters and axion stars as a function of the parameters $m_a$ and $n$. 
They found that the fraction of DM contained in ASs $f_\star\ll 1$ is generally much lower than previously assumed, but confirmed that a significant fraction of the galactic AS cores can reach masses close to $M_{\star,\lambda}$ - even from linear growth predictions alone.
Importantly, the mass distributions of axion stars and miniclusters drift to larger masses through repeated merger events over time, with the maximum MC mass reaching to larger $\mathcal{M}$ for larger $n$, as we will see later. 

Once an axion star has formed inside its host minicluster, it can grow in mass through accretion of axions from the (minicluster) background, which is why the growth rate will depend on the density and mass of its host MC. 
Accordingly, minicluster characteristics will determine both the axion star properties and the corresponding accretion rates. 
The characteristic minicluster mass $\mathcal{M}_0$ is calculated from the oscillation temperature $T_\mathrm{osc}$ for different axion models according to
\begin{align}
    \mathcal{M}_0 = \rho_{a,0} \frac{4\pi}{3}\left(\frac{\pi}{a_{\rm osc} H_{\rm osc}}\right)^3
    \label{eq:M0}
\end{align}
with $a_{\rm osc}\equiv a(T_\mathrm{osc})$ and $H_{\rm osc}\equiv H(T_\mathrm{osc})$ \cite{Fairbairn_2018}.
Note that $\mathcal{M}_0\propto T_{\rm osc}^{-3}$. Thus for fixed $m_a$, smaller $n$ implies that the axion mass becomes relevant earlier, i.e., larger $T_\mathrm{osc}$, which leads 
\footnote{That is for our approach of setting $f_a$ by fixing the relic abundance in Eq.~\eqref{eq:Omega_a} to $\Omega_ah^2=0.12$.} 
to larger $\mathcal{M}_0$ (see Ref.~\cite{Maseizik:2024qly} for details).

We allow for variation of the present-day minicluster density
\begin{align}
\rho_{\rm mc} &\simeq 7\times 10^6 \, \Phi^3 (1+\Phi)
\,\mathrm{GeV}/\mathrm{cm}^{3} \,,
\label{eq:rho_mc}
\end{align}
by considering inital overdensity parameters $\Phi \in [0,10^4]$ following the prediction from Ref.~\cite{Kolb_1994} and where the distribution of $\Phi$ follows the fit in Eq.~\eqref{eq:PDF} and Fig.~\ref{fig:CDF_Phi}.
The range of minicluster masses defining the range of axion stars in our analysis is determined by the parametrization from Fairbairn et al. \cite{Fairbairn_2018}, who introduced the low- and high-mass cutoffs
\begin{align}
\mathcal{M}_{\min}(m_a,n) \Big |_{z=0} &\simeq \mathcal{M}_0(m_a,n) / 25, \label{eq:M0_cutoff}
\\
\mathcal{M}_{\max}(m_a,n) \Big |_{z=0} &\simeq 4.9 \times 10^6 \mathcal{M}_0(m_a,n) \label{eq:M_h_max} .
\end{align}
of the minicluster mass distribution at present-day redshift $z=0$.

Note that by applying the $\mathcal{M}_0$-cutoff in Eq.~\eqref{eq:M0_cutoff} we have neglected the low-mass component of the minicluster mass function (c.f. the different low-mass cutoffs of the MCMF discussed in Ref.~\cite{Maseizik:2024qly}).
Typical low-mass MCs with $\Phi\sim 1$ and $\mathcal{M}<\mathcal{M}_0$ have $M_\star \ll M_{\star,\lambda}$ and are thus unlikely to reach the critical AS mass within the timescales we are interested in.
Additionally, their long-time survival is uncertain due to tidal disruption in the galactic environment \cite{Kavanagh:2020gcy, Shen_2022} and energy-loss from repeated axion bursts, especially for the largest $\Phi\sim 10^4$.

In the range $\mathcal{M}_{\min} \leq \mathcal{M} \leq \mathcal{M}_{\max}$ the MCMF can be parametrized in the form
\begin{align} \label{eq:dn_dlnM}
    \frac{dn}{d\ln \mathcal{M}}(\mathbf{r}) = C_{\rm n}(\mathbf{r})
    \left(
    \frac{ \mathcal{M}}{ \mathcal{M}_0}
    \right)^{-1/2}
    \,,
\end{align}
where again we use the analytical prediction by \cite{Fairbairn_2018}.
The normalization constant $ C_{\rm n}(\mathbf{r})$ is fixed by normalization of the total MC mass to the mass of the Navarro-Frenk-White (NFW) DM halo of the Milky Way (see \cite{Maseizik:2024qly} for details).
Following our previous work \cite{Maseizik:2024qly}, we use the core-halo relation by Schive et al. \cite{Schive:2014dra} evaluated at matter radiation equality redshift $z\simeq z_\mathrm{eq}\simeq 3402$ to obtain the distribution of axion stars from the MCMF:
\begin{align}
M_\star(z) & 
=  \mathcal{M}_{h,\min}(z) \left(\frac{\mathcal{M}}{\mathcal{M}_{h,\min}(z)}\right)^{1 / 3}  ,\label{eq:CoreHalo}
\end{align}
where the redshift-dependent minimum halo mass
\begin{widetext}
\begin{align} \label{eq:M_h_min_CoreHalo}
\mathcal{M}_{h,\min} &= 8.34 \times 10^{-14}  M_\odot \, 
\left( \frac{1+z}{1+z_\mathrm{eq}} \right)^{3/4} \left(\frac{\zeta(z)}{\zeta(z_\mathrm{eq})}\right)^{1 / 4} \left(\frac{m_a}{\mu\text{eV}}\right)^{-3/2}\, 
\end{align}
\end{widetext}
can be interpreted as the minimum mass above which a minicluster can host an axion star (sometimes referred to as a solitonic core).
\footnote{
It should be emphasized that the gravitational core-halo relation \eqref{eq:CoreHalo} remains valid for the dilute stable AS configurations and weak attractive self-interactions $(\lambda=-m_a^2/f_a^2)$ considered in this work.
For the case of dense axion stars and for strong self-couplings, the dominance of gravity over self-interactions would be violated, rendering the core-halo relation \eqref{eq:CoreHalo} inadequate.
}
Note that as opposed to $\mathcal{M}_{\min}$ from Eq.~\eqref{eq:M0_cutoff}, the mass cutoff \eqref{eq:M_h_min_CoreHalo} only applies to the formation of axion stars and that miniclusters with $\mathcal{M}< \mathcal{M}_{h,\min}$ can still form without a soliton core.

We calculate the accretion rate of ASs obtained from the self-similar attractor model in \cite{Dmitriev:2023ipv} by means of the condensation time
\begin{align} \label{eq:tau_gr}
\tau_{\rm gr} \simeq \frac{5.7\times 10^6 \,\mathrm{yr}}{\Phi^3(1+\Phi)}\left(\frac{\mathcal{M}}{10^{-12} M_{\odot}}\right)^2\left(\frac{m_a}{\mu \mathrm{eV}}\right)^3
,
\end{align}
which constitutes a characteristic timescale for the self-gravitating AS-MC system \cite{Levkov:2018kau}.
The corresponding accretion rate of the self-similar attractor is \cite{maseizik_seong_2024, Dmitriev:2023ipv}
\begin{widetext}
\begin{align} \label{eq:acc_Levkov}
    \frac{\delta M_\star}{\delta t}
    \simeq
    \frac{ \displaystyle
    \left(1-\frac{M_\star}{\mathcal{M}}\right)^6}
    {\displaystyle
    \left[5+\frac{1}{\epsilon^2}\left(\frac{M_\star}{\mathcal{M}}\right)^2\left(9-4\frac{M_\star}{\mathcal{M}}
    \right)\right]\left[1+\frac{1}{\epsilon^2}\left(\frac{M_\star}{\mathcal{M}}\right)^3\right]^2}
    \frac{\mathcal{M}}{1.1\,\tau_{\rm gr}}
    \,,
\end{align}
\end{widetext}
where $\epsilon$ is determined by \cite{Dmitriev:2023ipv}
\begin{align}
    \epsilon \simeq 0.086 \sqrt{\Phi} (1+\Phi)^{1/6} \left( \frac{10^{-12}\, M_\odot}{\mathcal{M}} \right)^{2/3} \left( \frac{\mu\text{eV}}{m_a} \right)
    .
    \label{eq:epsilon}
\end{align}
This means that for a given axion mass, the accretion rate generally depends on three parameters: the minicluster density parameter $\Phi$, the MC mass $\mathcal{M}$ and the axion star mass $M_\star$.
We assume the two parameters $\mathcal{M}$ and $\Phi$ to be independent from each other and from the galactocentric radial coordinate $R$ for simplicity.
This allows us to integrate over $\Phi$ in the range $[0,10^4]$ where the MC overdensity parameter follows the probability distribution $p_\Phi(\Phi)$ from the Pearson-VII fit in Eq.~\eqref{eq:PDF}.
Similarly, the minicluster masses in our consideration follow the MCMF from our previous work \cite{Maseizik:2024qly} for the $\mathcal{M}_0$-cutoffs given by Eqs.~\eqref{eq:M0_cutoff} and \eqref{eq:M_h_max}.
Note that by taking the MCMF from \cite{Maseizik:2024qly}, we have implicitly applied several consistency cutoffs.
The most important of the latter is the consideration of the minimum MC mass $\mathcal{M}_{h,\min}$ from Eq.~\eqref{eq:M_h_min_CoreHalo} ensuring that $M_\star \leq \mathcal{M}$ (see Ref.~\cite{Maseizik:2024qly} for details).

We estimate the number of bosenovae by integrating over the range of galactic MCs with properties $\{\mathcal{M},\Phi\}$ and determine how many of these systems can exceed the critical AS mass through self-similar accretion onto the soliton core in a given time.
An important observation here is the fact that the accretion rates obtained from Eq.~\eqref{eq:acc_Levkov} decrease monotonically with increasing $M_\star$ for a given $\mathcal{M}$ as shown in Fig.~\ref{fig:AccLevkov}.
Moreover, the high-mass tail of the MCMF with $\mathcal{M}> \mathcal{M}_0$ provides a major contribution to the number of MCs hosting an AS core close to the critical mass $M_\star \sim M_{\star,\lambda}$.
Due to the scaling $M_\star \propto \mathcal{M}^{1/3}$ of the core-halo relation in Eq.~\eqref{eq:CoreHalo}, these near-critical AS-MC systems exhibit very small values of $M_\star/\mathcal{M}$, which renders the axion star accretion rate $\delta M_\star / \delta t \propto \mathcal{M}/\tau_{\rm gr}\propto 1/\mathcal{M}$ in Eq.~\eqref{eq:acc_Levkov} smaller for larger $\mathcal{M}$.
This leads us to take the simplifying but conservative assumption that every axion star accretes with a rate similar to that of a critical AS system, i.e. $\delta M_\star / \delta t \sim \delta M_\star(\mathcal{M}_\lambda, M_{\star,\lambda}, \Phi) / \delta t$.

Under this assumption, we can extrapolate the minimum AS mass in the ASMF, which reaches criticality over an observation time $t_\mathrm{obs}$ as
 \begin{align} \label{eq:M_star_acc}
   M_{\star,\mathrm{acc}}(\Phi, t_\mathrm{obs}) &= M_{\star,\lambda} - \frac{\delta M_\star(\mathcal{M}_\lambda, M_{\star,\lambda}, \Phi)}{\delta t}\, t_\mathrm{obs},
\end{align}
again evaluated at the critical mass $M_{\star,\lambda}$ and corresponding MC mass $\mathcal{M}_\lambda$ for simplicity.
We estimate the MC threshold mass corresponding to the AS mass $M_{\star,\mathrm{acc}}$ from the core-halo relation, i.e.
\begin{align} \label{eq:M_acc}
    \mathcal{M}_{\lambda,\mathrm{acc}} &\equiv \mathcal{M} (M_{\star, \mathrm{acc}}) \,.
 \end{align}
Any axion star with initial mass $M_\star \geq M_{\star,\mathrm{acc}}(t_\mathrm{obs})$ will accrete enough axion dark matter from its surrounding minicluster within a given time $t_\mathrm{obs}$ to become super-critical.
The number of the resulting bosenovae expected from these super-critical AS-MC systems can then be calculated from the MCMF according to
\begin{widetext}
\begin{align} \label{eq:N_Novae}
\mathcal{N}_\mathrm{Nova}(t_\mathrm{obs}) 
= 4 \pi \int d R\, R^2 \int d \Phi\, p_{\Phi}(\Phi) \mathcal{P}_{\mathrm{surv}}(\Phi)
\int_{\mathcal{M}_{\lambda,\text{min}}(\Phi, t_\mathrm{obs})}^{\mathcal{M}_{\lambda,\max}} d \mathcal{M}\, \frac{dn}{d\mathcal{M}}(R) \,,
\end{align}
\end{widetext}
where $\mathcal{M}_{\lambda,\max}=\min(\mathcal{M}_{\max},\mathcal{M}_\lambda)$, and $\mathcal{M}_{\lambda,\mathrm{min}}(\Phi, t_\mathrm{obs})$ is an effective low-mass cutoff derived from $\mathcal{M}_{\lambda,\mathrm{acc}}$ (see App. \ref{app:Cutoffs} and Eq.~\eqref{eq:M_lambda_min}).
Here $\mathcal{P}_\mathrm{surv}(\Phi)$ accounts for the survival rate of miniclusters in the stellar environment due to tidal stripping as defined by Eq.~\eqref{eq:P_surv} from Ref.~\cite{Shen_2022} and $dn/d\mathcal{M}$ is the MC number distribution obtained from the MCMF in Eq.~\eqref{eq:dn_dlnM}.
From the scaling $\delta M_\star \propto \tau_\mathrm{gr}^{-1} \propto \Phi^4$ in Eq.~\eqref{eq:acc_Levkov}, we can already see that the strongest contribution to $\mathcal{N}_\mathrm{Nova}$ is given by the densest miniclusters, which have $\Phi \lesssim 10^4$.
In some extreme cases with $\Phi \approx 10^4$, the accretion rates can become large enough for $M_{\star,\mathrm{acc}}$ in Eq.~\eqref{eq:M_star_acc} to reach negative values, when $\delta M_\star / \delta t \cdot t_\mathrm{obs} > M_{\star,\lambda}$.
We drop the corresponding AS-MC population with $\{\mathcal{M},\Phi\}$ predicting $M_{\star,\mathrm{acc}}<0$, due to uncertainties about their long-time stability.

For the QCD axion with $n=3.34$, $m_a=50\,\mu$eV and critical AS/MC masses $M_{\star,\lambda}\simeq 2\times 10^{-13} \,M_\odot$ and $\mathcal{M}_\lambda\simeq 2 \times 10^{-7} \,M_\odot$, we find that typical systems with $\Phi\simeq 1$ should have $\delta M_\star(\mathcal{M}_\lambda, M_{\star,\lambda}, \Phi) / \delta t \simeq 2 \times 10^{-38}\, M_\odot\,$s$^{-1}$, which implies $M_{\star,\mathrm{acc}} \approx M_{\star,\lambda}$.
On the other hand, for the densest miniclusters with $\Phi \simeq 10^4$ we obtain $\delta M_\star(\mathcal{M}_\lambda, M_{\star,\lambda}, \Phi) / \delta t \simeq 10^{-22}\, M_\odot\,$s$^{-1}$, which gives $\delta M_\star/ \delta t \cdot t_\mathrm{obs} \simeq 4 \times 10^{-15}\, M_\odot$ for the accreted mass in Eq.~\eqref{eq:M_acc} after $t_\mathrm{obs}=1\,$yr, corresponding to an order one percent mass growth.

We also emphasize that by taking the ASMF obtained from the MCMF with the core-halo relation \eqref{eq:CoreHalo}, we are neglecting the long-term mass growth from the host MC onto its AS core, that was suggested by the simulations in Ref.~\cite{Dmitriev:2023ipv}.
Our approach thus constitutes a conservative estimate of the present-day ASMF, which does not account for possible time-dependence of the core-halo relation.
Future work can improve on this estimate by incorporating predictions of long-time AS growth, and possible modifications of the late-time core-halo relation for $t\gg \tau_\mathrm{gr}$ similar to what was done in Ref.~\cite{Dmitriev:2023ipv}.
Note however that such modeling would also provide large numbers of ASs with predicted masses $M_\star > M_{\star,\lambda}$, which we neglect due to large uncertainties in their detailed evolution.

\begin{figure}[t]
\centering
\includegraphics[width=\columnwidth]{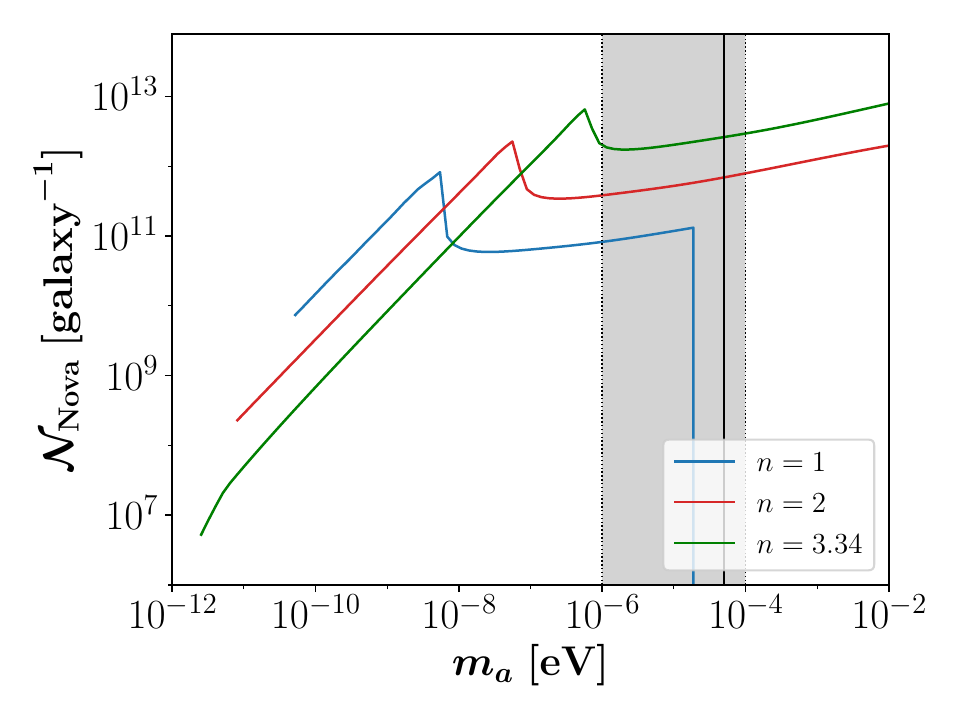}
\caption{Number of bosenovae occurring within the Milky Way halo for $t_\mathrm{obs}=1\,$yr for the $\mathcal{M}_0$-Cutoff, i.e. without the low-$\mathcal{M}$ tail of the MCMF.
Colors denote axion models with different temperature evolution according to Eq.~\eqref{eq:m_T}.
The grey-shaded region denotes the cosmological QCD axion mass band with $10^{-6}\,$eV$\leq m_a \leq 10^{-4}\,$eV and the black solid vertical indicates the QCD axion mass $m_a\approx 50\,\mu$eV assumed in this work.}
\label{fig:N_Novae}
\end{figure}

We illustrate the total number of galactic bosenovae derived from Eq.~\eqref{eq:N_Novae} for axion masses in the range $10^{-12}\,{\rm eV}\leq m_a \leq 10^{-2}\,{\rm eV}$ and for three axion models with different temperature evolution $n=1,2,3.34$ in Fig.~\ref{fig:N_Novae}.
The turn-around arises from the $m_a$-dependence of the accretion-induced low-mass cutoff $\mathcal{M}_{\lambda,\mathrm{acc}}$ derived from Eq.~\eqref{eq:M_star_acc} and from the cutoff-dependence of $\mathcal{M}_{\lambda,\min}$, $\mathcal{M}_{\lambda,\max}$ following Eqs.~\eqref{eq:M_lambda_min} and \eqref{eq:M_lambda_max}.
Notably, the number of galactic Bosenovae increases with larger $n$.

The different scaling of the peaks in Fig.~\ref{fig:N_Novae} is related to two competing effects: First, the increased number of ASs $\mathcal{N}_{\star,\mathrm{tot}} \propto 1 / \mathcal{M}_0$ \cite{Maseizik:2024qly} for smaller $n$ and $\mathcal{M}_0$, and secondly the increased accretion rates for larger $n$.
The first of these effects is a direct consequence of the normalization of the MCMF, which is set to match the total DM mass of the Milky Way, hence $\mathcal{N}_\mathrm{tot}\propto 1/\mathcal{M}_0$, where $\mathcal{N}_\mathrm{tot}$ is the total MC number.
On the other hand, the second effect is related to the scaling of the critical mass $M_{\star,\lambda} \propto f_a$, which inherits a temperature dependence from the decay constant $f_a$ fixed by the $n$-dependent relic abundance in Eq.~\eqref{eq:Omega_a}.
For larger $n$, this yields smaller critical AS masses $M_{\star,\lambda}$, which turn out to boost the accretion rates of the self-similar attractor in Eq.~\eqref{eq:acc_Levkov}.
As can be seen by the scaling of $\mathcal{N}_\mathrm{Nova}$ with larger $n$ in Fig.~\ref{fig:N_Novae}, the benefit of having larger accretion rates and smaller $M_{\star,\lambda}$ is dominant over the scaling of $\mathcal{N}_\mathrm{tot}$.

Lastly and in the case of small $n=1$ in blue, a sudden drop in $\mathcal{N}_\mathrm{Nova}$ arises from the different scalings of $\mathcal{M}_{\max}$ and $\mathcal{M}_\lambda$.
Eventually at some $m_a$, the accretion-induced critical mass $\mathcal{M}_{\lambda,\mathrm{acc}} > \mathcal{M}_{\max}$ lies beyond the range of the initial ASMF, yielding $\mathcal{N}_\mathrm{Nova}=0$.

We also mention for completeness, that the results in Fig.~\ref{fig:N_Novae} imply that the number of bosenovae occuring within a Hubble time could be as large as $\mathcal{N}_\mathrm{Nova} \, t_H \gg \mathcal{N}_\mathrm{tot}$ for some $(m_a,n)$.
This observation indicates that a large number of axion stars are expected to collapse repeatedly on cosmological timescales.
In fact, we can consider the exemplary case of the QCD axion with $m_a=50\,\mu$eV and $n=3.34$ to find that $M_{\star,\lambda} / \mathcal{M}_\lambda \sim 10^{-6}$, which means that a typical AS-MC system with near-critical AS/MC-masses can undergo  $\sim 10^6$ bosenovae until it is depleted of its total mass.
Answering the question of how many of the AS-MC systems are expected to shed their initial mass within $t_H$ requires investigation of the full time evolution of the MCMF, ASMF and core-halo relation - all of which are beyond the scope of this work.
See also Ref.~\cite{Fox_2023} for a similar study, which constrains axion models through depletion of cold dark matter following repeated bosenovae in the cosmological context.
Our analysis complements the work in Ref.~\cite{Fox_2023} by using the direct observation of relativistic axions from bosenovae in DM detectors.

\section{Signals from nearby bosenovae}
\label{sec:signals}
We are now in a position to determine the total number of galactic bosenova, which occur within a given observation time $t_\mathrm{obs}$, from Eq.~\eqref{eq:N_Novae}.
The next step to connect our AS accretion model to astrophysical observations is to estimate how many of the predicted axion bursts can actually be detected by existing and future DM experiments.
We provide an answer to this question by recalling the most important aspects of axion burst propagation and of the observed bosenova signal from Ref.~\cite{Eby:2021ece} in the following.

For a given detector, there will be a minimum distance at which a bosenova will be detectable on Earth. Following~\cite{Eby:2021ece}, we characterize the detectability of a transient bosenova event by a ratio of the sensitivity to the bosenova relative to the sensitivity for a non-relativistic dark matter search. 
We first consider an interaction Lagrangian which is linear in $a$ with no derivative.
The canonical example is the axion-photon interaction
\begin{equation} \label{eq:L_gamma}
    \Lcal \supset g a F^{\mu\nu}\tilde{F}_{\mu\nu}\,,
\end{equation}
where $g$ is the axion-photon coupling and $F^{\mu\nu}$ the electromagnetic field strength tensor, with $\tilde{F}^{\mu\nu}$ its dual.
For different axion models $(m_a,n)$, we set $g=c/f_a$, where the model-dependent coefficient $c$ may be chosen arbitrarily, except in the case of the QCD axion with $m_a=50\,\mu eV$, $n=3.34$ and $c=\alpha(E/N - 1.92)/(2\pi)$, where $E,N$ are the electromagnetic and color anomaly numbers and $\alpha$ is the fine-structure constant.

In a given experiment, if the minimal detectable coupling of non-relativistic DM is $g_{\rm dm}$ and the minimal detectable bosenova coupling is $g_\star$, we have \cite{Eby:2021ece}
\begin{equation} \label{eq:gratio}
    \frac{g_\star}{g_{\rm dm}} \simeq 
        \left(\frac{\rho_{\rm dm}}{\rho_\star}
        \right)^{1/2}
        \frac{t_{\rm obs}^{1/4}{\rm min}[t_{\rm obs}^{1/4},\tau_{\rm dm}^{1/4}]}
        {{\rm min}[t_{\rm obs}^{1/4},\delta t^{1/4}]
        {\rm min}[t_{\rm obs}^{1/4},\tau_{\star}^{1/4}]}\,,
\end{equation}
where $\tau_{\rm dm} \simeq 2\pi/(m_a v_{\rm dm}^2)$ is the DM coherence time (see e.g.~\cite{Budker:2013hfa}), $\delta t$ and $\tau_\star$ are the duration and coherence time (respectively) of the relativistic burst at the position of the detector, and $t_{\rm obs}$ is the observation time of the experiment, which we take to be of order $1~{\rm year}$ for this study. 
In the presence of tidal streams, $\rho_\mathrm{dm}$ is close to the canonical value $\rho_\mathrm{dm}=0.4\,$GeV/cm$^3$.
Without tidal streams, the local DM density is a factor of 4 smaller, arising from the MC DM abundance of $f_\mathrm{mc}=0.75$, which we assume following Refs.~\cite{Maseizik:2024qly, Eggemeier:2019jsu}, giving $\rho_\mathrm{dm}=(1-f_\mathrm{mc})\,0.4\,$GeV/cm$^3$ for the local DM background.
According to Eq.~\eqref{eq:gratio}, this would change $g_\star/g_\mathrm{dm}$ by a factor of $1/2$ thus improving the detectability of Bosenovae compared to the cold DM case.

Note that if the DM-SM coupling is instead quadratic, then the dependence on the energy density in Eq.~\eqref{eq:gratio} is steeper, proportional to $(\rho_{\rm dm}/\rho_\star)$, leading to an enhanced sensitivity when $\rho_\star \gg \rho_{\rm dm}$ relative to the linear case; we will not consider this case in this work (see e.g.~\cite{Arakawa:2023gyq,Arakawa:2024lqr} for discussion).

For linear couplings which contain a derivative of the axion field, $\propto \nabla a$, the experimental sensitivity to $g$ gains an additional factor of the speed of the field, $g\propto v^{-1}$. 
The canonical example here is the axion-fermion coupling
\begin{equation} \label{eq:L_aff}
    \Lcal \supset g_\nabla (\partial_\mu a) \bar{\psi} \gamma^\mu \gamma^5 \psi\,,  
\end{equation}
where $\psi$ is a SM fermion field.
This factor suppresses the sensitivity by $v_{\rm dm}^{-1} \sim 10^3$ in the non-relativistic DM case, giving searches for relativistic fields a comparative advantage. The corresponding sensitivity ratio with derivative couplings takes the form \cite{Eby:2021ece}
\begin{equation} \label{eq:gratio_der}
    \left(\frac{g_\star}{g_{\rm dm}}\right)_\nabla \simeq  v_{\rm dm}
        \left(\frac{\rho_{\rm dm}}{\rho_\star}
        \right)^{1/2}
        \frac{t_{\rm obs}^{1/4}{\rm min}[t_{\rm obs}^{1/4},\tau_{\rm dm}^{1/4}]}
        {{\rm min}[t_{\rm obs}^{1/4},\delta t^{1/4}]
        {\rm min}[t_{\rm obs}^{1/4},\tau_{\star}^{1/4}]}\,.
\end{equation}
As we will see in the next section, this factor of $10^3$ enhancement in sensitivity ratio motivates ongoing and future experiments searching for axions with derivative couplings, e.g. CASPEr~\cite{Budker:2013hfa,JacksonKimball:2017elr} which is already underway but designed for resonant searches.
Note also that since $g_\star / g_{\rm dm}$ is a ratio of sensitivities, the results in Eqs.~\eqref{eq:gratio} and \eqref{eq:gratio_der} are independent of the properties of a particular broadband-type experiment.

Irrespective of the nature of the SM coupling, the observed bosenova energy density is simply given by
\begin{equation}
    \rho_\star \simeq \frac{\Ecal}{4\pi d_\mathrm{obs}^2 \delta x}\,,
\end{equation}
where $d_\mathrm{obs}$ is the distance to the bosenova, $\Ecal$ is the total energy emitted from the source, and $\delta x \simeq \delta t$ for relativistic particles.
Notably, Ref.~\cite{Eby:2021ece} found that due to wave spreading in flight\footnote{This approximation is justified as long as the intrinsic burst duration is negligible compared to the actual duration; for bosenovae a distance $d_\mathrm{obs}\simeq {\rm pc}$ (kpc) away, this is true when $m_a\gtrsim 10^{-16}\,{\rm eV}$ ($m_a\gtrsim 10^{-19}\,{\rm eV}$), which easily holds for the parameters considered in this work. See~\cite{Eby:2021ece} for details.}, the duration and coherence time of the burst at the position of the detector grow with $d_\mathrm{obs}$ as
\begin{equation} \label{eq:deltat_taustar}
    \delta x \simeq \delta t \simeq \frac{\delta k_a}{m_a}\frac{d_\mathrm{obs}}{q^3}, \qquad 
    \tau_\star \simeq \frac{\pi d_\mathrm{obs}}{200 q^3}\,,
\end{equation}
where $k_0 \equiv q m_a$ is the peak momentum and $\delta k_a$ is the momentum spread, and we have assumed $q^2+1\simeq q^2$. 
For the leading relativistic momentum peak in the bosenova spectrum, $q\simeq 3$ and $\delta k_a \simeq m_a$, and the total energy emitted is $\Ecal \simeq f_{\rm em} M_{\star,\lambda}$ with $f_{\rm em} \simeq 0.2-0.5$~\cite{Levkov:2016rkk}, where we assume $f_{\rm em} = 0.3$ in the following.
We see from Eq.~\eqref{eq:deltat_taustar} that for a bosenova, $\delta t \simeq 10^{-2}d_{\rm obs}$, implying a long duration of the signal at the position of the detector. 
As mentioned in Ref.~\cite{Eby:2021ece}, the spread in momentum is much larger than the cold DM case (where $\delta k_a \simeq 10^{-3} m_a$), implying that broadband searches are more well-suited to detecting bosenovae compared to resonant-type searches. 

Summarizing the above, for a given axion model $(m_a,n)$, and fixing the search timescale $t_{\rm obs}\simeq 1\,{\rm year}$, Eq.~\eqref{eq:gratio} is a function of $d_\mathrm{obs}$ only. Therefore, for a given choice of input parameters, we can estimate the maximal distance $d_{\rm max}$ of a detectable bosenova, defined by the value of $d_{\rm obs}$ which gives $g_\star/g_{\rm dm}(d_\mathrm{obs})=1$. We will see below that this maximal distance is typically a few orders of magnitude smaller than a parsec for non-derivative couplings, and somewhat larger for derivative couplings, depending on the input parameters.
This means that for observation distances $d_\mathrm{obs} \leq d_{\max}$, or equivalently for sensitivity ratios $g_\star/g_{\rm dm} \leq 1$, bosenovae can be detected with different types of broadband axion DM experiments \cite{Eby:2021ece}.

Since the maximum observable distance derived from Eq.~\eqref{eq:gratio} is small $d_\mathrm{max}\lesssim 1\,$pc compared to galactic length scales $\sim 1\,$kpc, we can estimate the typical distance between two bosenovae by writing
\begin{align} \label{eq:d_avg}
    \langle d \rangle &\sim \left( \frac{\text{Volume}~V}{\mathcal{N}_\mathrm{Nova}|_{\text{within}~V}} \right)^{1/3}
    \sim \left( \frac{4 \pi R_\odot^3/3}{f_\odot \mathcal{N}_\mathrm{Nova}} \right)^{1/3} \,,
\end{align}
where $R_\odot=8.3\,$kpc is the solar position and $f_\odot\approx 0.032$ is the fraction of MCs contained within $r\leq R_\odot$.
Physically, Eq.~\eqref{eq:d_avg} gives the average distance between $f_\odot \mathcal{N}_\mathrm{Nova}$ events within a spherical volume of radius $R_\odot$, where the NFW distributed nature of the events is accounted for by $\mathcal{N}_\mathrm{Nova}$ following Eq.~\eqref{eq:N_Novae}.
From this we define the average observation distance as 
\begin{align} \label{eq:d_obs}
    \langle d_\mathrm{obs} \rangle = \frac{\langle d \rangle}{2}
\end{align}
differing only by the geometrical factor of $1/2$.
We emphasize that in our framework and for each axion model $(m_a,n)$, the AS properties derived from the MCMF following Ref.~\cite{Maseizik:2024qly} as well as the accretion modelling based on Eqs.~\eqref{eq:M_star_acc} and \eqref{eq:N_Novae} yield a fixed average observation distance $\langle d_\mathrm{obs} \rangle$, which needs to be compared to the maximum observable distance $d_{\max}$.
As we will show in the following, the sensitivity ratios of these experiments, given by Eqs.~\eqref{eq:gratio} and \eqref{eq:gratio_der}, depend strongly on the axion model and coupling.

\subsection{Non-Derivative Coupling}
\begin{figure}[t]
\centering
\includegraphics[width=\columnwidth]{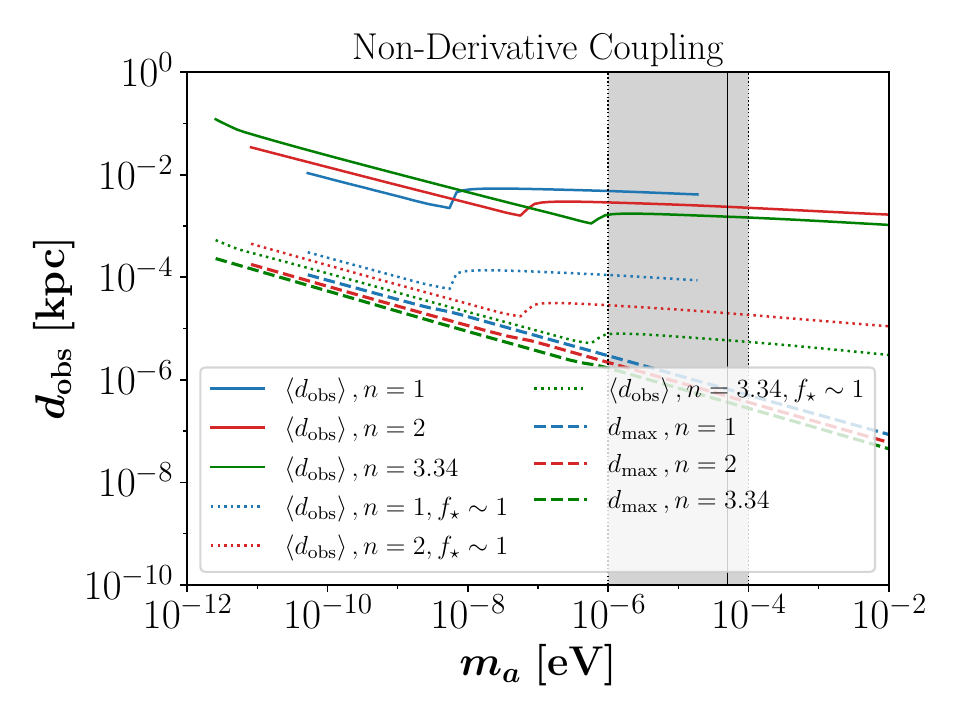}
\caption{Typical observation distance $\langle d_\mathrm{obs} \rangle$ of galactic bosenovae in solid colored lines, the same for the case that $f_\star\sim 1$ in dotted lines, and maximum observable distance $d_{\max}$ for $g_\star/g_{\rm dm}=1$ in dashed lines.
The colors denote axion models with different temperature evolution according to Eq.~\eqref{eq:m_T}.
Shown for axions with non-derivative couplings using the $\mathcal{M}_0$-Cutoff of the MCMF \cite{Maseizik:2024qly}.
The grey-shaded region denotes the cosmological QCD axion mass band with $10^{-6}\,$eV$\leq m_a \leq 10^{-4}\,$eV and the black solid vertical line indicates the QCD axion mass $m_a\approx 50\,\mu$eV assumed in this work.}\label{fig:d_obs_projected}
\end{figure}

We start our analysis for the case of axions with non-derivative couplings, e.g., the axion-photon interaction from Eq.~\eqref{eq:L_gamma}, in Fig.~\ref{fig:d_obs_projected}.
We plot the contour lines where $g_\star/g_\mathrm{dm}(d_\mathrm{obs})=1$ and the axion bursts emitted from collapsing ASs are sufficiently close to be distinguishable from the background DM.
The contour lines thus represent the maximum observable distance $d_{\max}$ of broadband axion DM experiments for bosenova events following Eq.~\eqref{eq:gratio}.
This maximum distance needs to be compared with the average observed distance of galactic bosenovae.
For the latter we use the average observed distance of collapsing AS cores $\langle d_\mathrm{obs} \rangle$ from Eq.~\eqref{eq:d_obs} as an estimate.
Following this approach, regions of $m_a$, where $\langle d_\mathrm{obs} \rangle \leq d_{\max}$ can be probed by current and upcoming broadband experiments.
Fig.~\ref{fig:d_obs_projected} thus demonstrates, that using the MCMF from \cite{Maseizik:2024qly} and for composite AS-MC systems obeying the core-halo relation \eqref{eq:CoreHalo}, bosenovae occur too rarely to be detected in axion DM searches.

Nevertheless, there are several considerations that can improve these predictions.
First and mainly, we have neglected long-time AS accretion, which is expected to significantly boost the number of bosenovae.
Secondly, the relative burst sensitivity of future telescopes could be improved, for example by including spectral information about the signal or by performing dedicated burst searches, thus enhancing the maximum observable distance $d_\mathrm{max}$ for bosenovae detection.
And lastly, there have been recent studies, namely Refs.~\cite{Gorghetto:2024vnp} and \cite{Chang:2024fol}, which suggest that an order-one fraction of the total galactic dark matter may be contained in the form of axion stars, rather than miniclusters as we have assumed.
We can characterize these models in our present analysis by setting the relative AS DM abundance
\begin{align}
    f_\star = \frac{M_\mathrm{\star,tot}}{\mathcal{M}_{200}},
\end{align}
determined from the total mass of galactic ASs $M_\mathrm{\star,tot}$ and from the mass $\mathcal{M}_{200}$ of the Milky Way DM halo, equal to 1.
Recalling the results from Ref.~\cite{Maseizik:2024qly}, our approach predicts $f_\star \sim 10^{-7}$ to $f_\star \sim 10^{-4}$, depending on $n$.
Setting $f_\star \sim 1$ would thus boost the total number of ASs (and thus also $\mathcal{N}_\mathrm{Nova}$) by a linear factor of $f_\star$.
According to Eq.~\eqref{eq:d_avg}, this can significantly lower the expected average distance of bosenovae.
The resulting reduction is of order $\sim 10^{-2}$ for the average distance $\langle d_\mathrm{obs} \rangle \propto f_\star^{-1/3}$.
We plot the predictions from setting $f_\star=1$ in dotted colored lines in Fig.~\ref{fig:d_obs_projected} and find that this assumption allows for detection of bosenovae with only minor improvements in the sensitivity ratio and for axion models with small $m_a <10^{-6}\,$eV.

Note that our predictions for $\langle d_\mathrm{obs} \rangle$ in Fig.~\ref{fig:d_obs_projected} are still relying on the initial ASMF from Ref.~\cite{Maseizik:2024qly} without the long-time AS mass growth predicted by Ref.~\cite{Dmitriev:2023ipv}.
Incorporating long-time AS accretion could lead to a pile-up of axion stars around $M_\star \approx M_{\star,\lambda}$ thus enhancing the expected number of galactic Bosenovae $\mathcal{N}_\mathrm{Nova}$, and possibly reducing the predicted values $\langle d_\mathrm{obs} \rangle$ below the threshold of observability $d_{\max}$.
Better understanding of AS mass growth could still yield observable signatures even in the case of axion-photon couplings.

\subsection{Derivative Coupling}
\begin{figure}[t]
\centering
\includegraphics[width=\columnwidth]{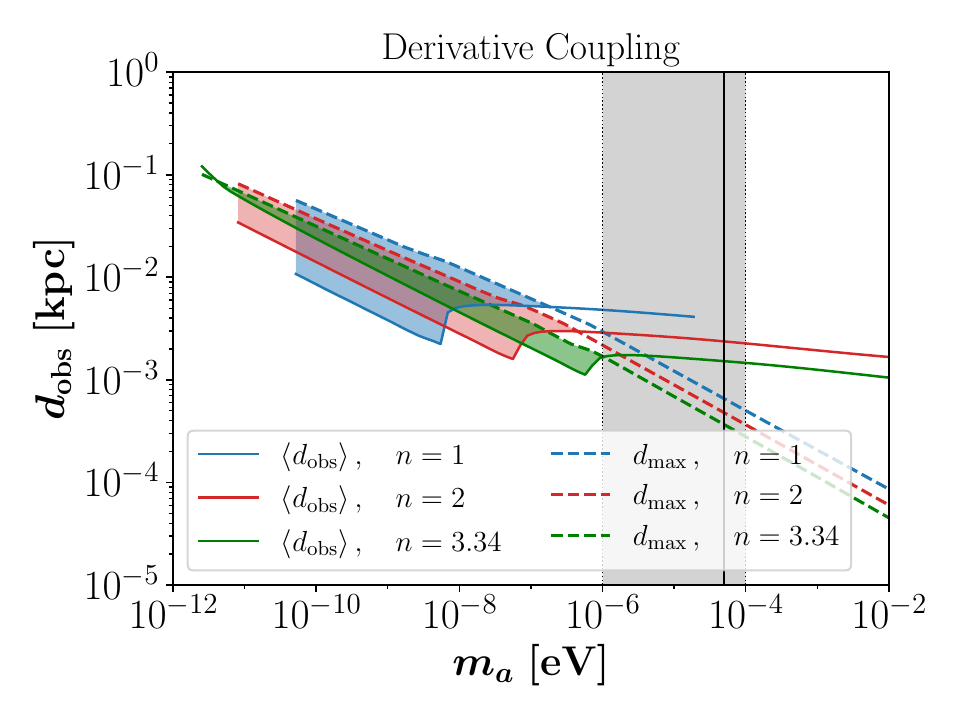}
\caption{Typical observation distance $\langle d_\mathrm{obs} \rangle$ of galactic bosenovae in solid colored lines and maximum observable distance $d_{\max}$ given by $(g_\star/g_\mathrm{dm})_{\nabla}=1$ in dashed lines.
Colors denote axion models with different temperature evolution according to Eq.~\eqref{eq:m_T}, calculated for axions with derivative couplings using the $\mathcal{M}_0$-Cutoff.
Color-shaded regions indicate the axion mass range, where bosenovae are detectable, i.e. $\langle d_\mathrm{obs} \rangle \leq d_\mathrm{max}$.}\label{fig:d_obs_der}
\end{figure}

The sensitivity ratio of DM search experiments exploiting derivative-type axion couplings like the axion-fermion interaction in Eq.~\eqref{eq:L_aff} is boosted by the ratio of non-relativistic to burst DM speeds, roughly $v_{\rm dm}/v_\star \simeq 10^{-3}$.
This circumstance increases the maximum observable distance $d_{\max}$ determined from setting $(g_\star/g_\mathrm{dm})_\nabla=1$ in Eq.~\eqref{eq:gratio_der}, as shown in Fig.~\ref{fig:d_obs_der} in dashed colored lines.
The resulting boost in relative burst sensitivity is sufficient to render a large part of the axion-mass range detectable.
Regions of $m_a$ with $\langle d_\mathrm{obs} \rangle \leq d_\mathrm{max}$ are highlighted by colored shades.

Depending on the temperature evolution $n$ of the axion mass, the occurrence of bosenovae is restricted by the existence of miniclusters, derived from the tensor-to-scalar ratio constraint in the post-inflationary scenario which truncates the high-$f_a$ range, or equivalently, low-$m_a$ range \cite{Fairbairn_2018, Maseizik:2024qly}.
The detailed scaling with $n$ of the solid lines in Fig.~\ref{fig:d_obs_der} arises from a combination of the scaling of the decay constant $f_a$ fixed by Eq.~\eqref{eq:Omega_a}, the scaling of the characteristic MC mass $\mathcal{M}_0$ from Eq.~\eqref{eq:M0} and the accretion rate \eqref{eq:acc_Levkov} used to determine $M_{\star,\mathrm{acc}}$ in Eq.~\eqref{eq:M_star_acc}.

\begin{figure}[t]
\centering
\includegraphics[width=\columnwidth]{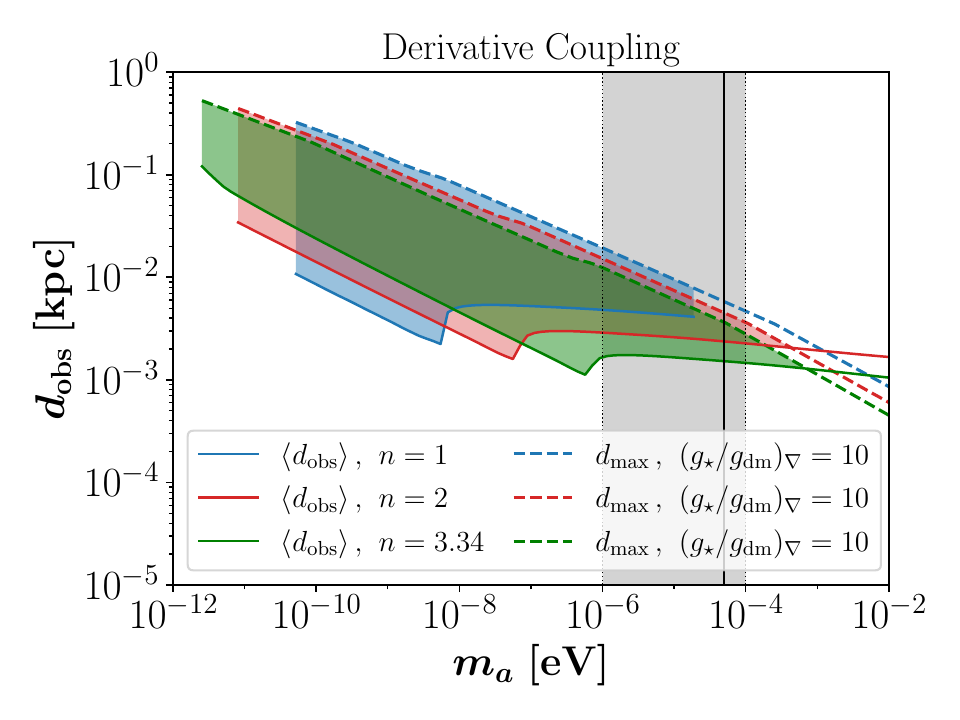}
\caption{Typical observation distance $\langle d_\mathrm{obs} \rangle$ of galactic bosenovae in solid colored lines and \textit{projected} maximum observable distance $d_{\max}$ of future experiments with $(g_\star/g_\mathrm{dm})_{\nabla}=10$ in dashed lines.
Shown for axions with derivative couplings using the $\mathcal{M}_0$-Cutoff.
Color-shaded regions indicate the axion mass range, where bosenovae are detectable, i.e. $\langle d_\mathrm{obs} \rangle \leq d_\mathrm{max}$.}\label{fig:d_obs_der_projected}
\end{figure}

Coincidentally, the cosmological axion band indicated by the grey-shaded regions in Fig.~\ref{fig:d_obs_der} is just beyond detectability for $d_{\max}$ set by $(g_\star/g_\mathrm{dm})_\nabla=1$.
In the future, dedicated Bosenova searches using information on the energy spectrum of the burst could be used to improve the sensitivity $g_\star$ in Eq.~\eqref{eq:gratio_der} relative to the cold DM sensitivity $g_\mathrm{dm}$ (see Ref.~\cite{Eby:2021ece} for discussion).
We can therefore estimate the prospects of axion burst DM searches with increased sensitivity ratios, assuming an improvement of order 10, which is equivalent to relaxing the condition for detectability to $(g_\star/g_\mathrm{dm})_\nabla=10$ as depicted in Fig.~\ref{fig:d_obs_der_projected}.

In this scenario, the resulting average observed distance $\langle d_\mathrm{obs} \rangle$ of galactic bosenovae is sufficiently reduced to allow probes of axion models in the cosmological axion band $10^{-6}\,$eV$\leq m_a \leq 10^{-4}\,$eV using axion bursts.
Conveniently, the case $n=3.34$ with QCD-like temperature dependence of $m_a$ in green lines and shades covers nearly the entire range of the axion mass.
Even the other models with $n=1,2$ can be probed for a wide range of axion masses.
This enhancement in the case of derivative coupling searches motivates further innovation and potentially dedicated searches for bosenovae from such axions.

For experimental searches, we also provide the estimated number of Bosenovae occurring within an observation volume $V_\mathrm{obs}=4\pi d_{\max}^3/3$ assuming a constant DM density over the volume $V_\odot=4\pi R_\odot^3/3$, given by
\begin{align}
        \mathcal{N}_\mathrm{obs} \sim b \frac{V_\mathrm{obs}}{V_\odot} f_\odot \mathcal{N}_\mathrm{Nova}
    =  b \left(\frac{d_{\max}}{R_\odot} \right)^3 f_\odot \mathcal{N}_\mathrm{Nova} \,,
\end{align}
where $b=6/\pi$ is an order one coefficient introduced to set $\mathcal{N}_\mathrm{obs}=1\,$galaxy$^{-1}$ when $\langle d_\mathrm{obs} \rangle = d_{\max}$ for consistency using our definition of $\langle d_\mathrm{obs} \rangle$ in Eq.~\eqref{eq:d_obs}.

\begin{figure}[t]
\centering
\includegraphics[width=\columnwidth]{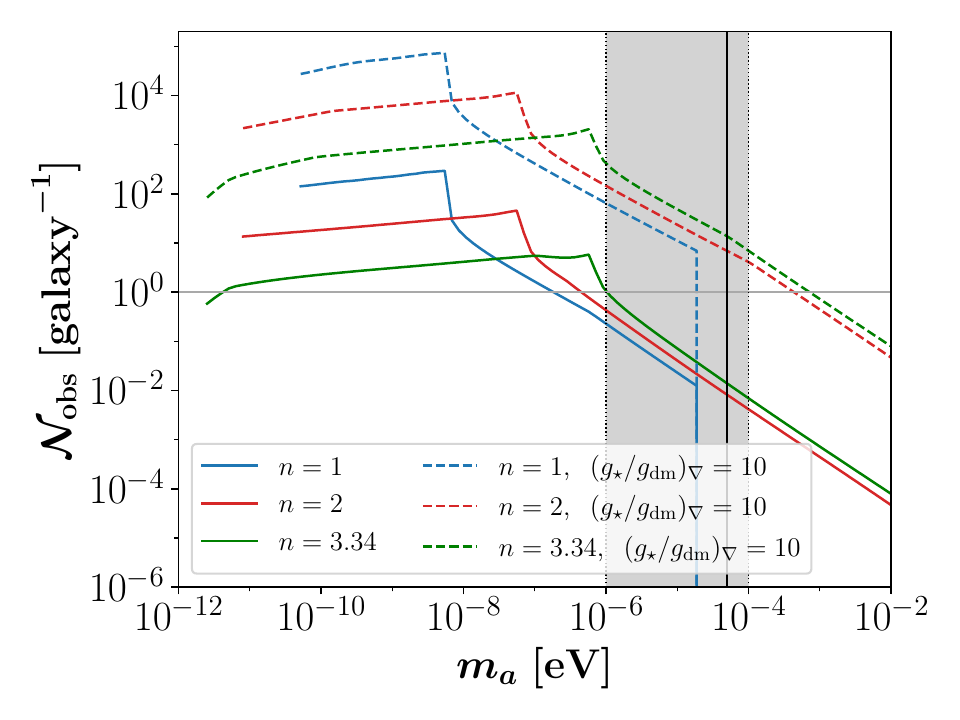}
\caption{Expected number of galactic bosenovae $\mathcal{N}_\mathrm{obs}$, occurring within a maximum observable distance $d_{\max}$ given by $(g_\star/g_\mathrm{dm})_{\nabla}=1$ in solid and $(g_\star/g_\mathrm{dm})_{\nabla}=10$ in dashed colored lines.
Shown for axions with derivative couplings using the $\mathcal{M}_0$-Cutoff and $t_\mathrm{obs}=1\,$yr.
The grey horizontal line indicates $\mathcal{N}_\mathrm{obs}=1\,$galaxy$^{-1}$.}\label{fig:N_Nova_obs_Der}
\end{figure}

We show the resulting number of observable Bosenovae $\mathcal{N}_\mathrm{obs}$ for $t_\mathrm{obs}=1\,$yr with $(g_\star / g_\mathrm{dm})_\nabla=1$ in solid and $(g_\star / g_\mathrm{dm})_\nabla=10$ in dashed colored lines in Fig.~\ref{fig:N_Nova_obs_Der}.
While a more sophisticated treatment of the detectability and galactic distribution of Bosenovae is required to give more concrete predictions, our rough estimations demonstrate that Bosenova signals are expected to occur even within smaller observation times than $t_\mathrm{obs}=1\,$yr.
In the case of order-one sensitivity ratios in solid lines, $\mathcal{N}_\mathrm{obs}$ ranges from order one to roughly $100$ observable events depending on ($m_a$, $n$), thus suggesting required observation times ranging from $t_\mathrm{obs}\sim 1\,$yr for $n=3.34$ down to $t_\mathrm{obs}\sim 3\,$d for $n=1$.

This requirement is further relaxed in the case of $(g_\star/g_\mathrm{dm})_{\nabla}=10$ in dashed lines, where all axion models $(m_a,n)$ allow $t_\mathrm{obs}\lesssim \mathcal{O}(10)\,$days for $m_a\lesssim 10^{-4}\,$eV.
The cosmological axion band and the canonical QCD axion parameters $m_a\approx 50\,\mu$eV are expected to yield $\mathcal{N}_\mathrm{obs}\sim 10$, which implies a required observation time of about a month.

\section{Discussion}
Our results suggest that for axion models with non-derivative couplings like the axion-photon coupling in Eq.~\eqref{eq:L_gamma}, bosenovae are unlikely to be detected in current axion DM searches.
Nevertheless, we have shown that with minor improvements in the relative burst sensitivity, our approach can be used to test predictions involving large AS DM abundance $f_\star \sim 1$ as suggested in Refs.~\cite{Gorghetto:2024vnp} and \cite{Chang:2024fol}.
For $f_\star \sim 1$ and different temperature evolution of $m_a$, axion masses with $m_a\lesssim 10^{-6}\,$eV could be probed in the future.
Importantly, our conservative treatment of the present-day ASMF from Ref.~\cite{Maseizik:2024qly} does not exclude the occurrence of bosenovae in experimental searches using the axion-photon coupling, since the number of burst signals can be further enhanced from considerations of long-time AS mass growth (see below).

Even without the boosted AS abundance and using conservative assumptions, we find that galactic bosenovae in axion models with derivative couplings, like the fermion coupling in Eq.~\eqref{eq:L_aff}, could be probed experimentally for a wide range of axion masses, ranging up to $10^{-12}\,$eV$\leq m_a \leq 10^{-6}\,$eV, depending on the temperature evolution $n$ of $m_a$.
Remarkably, our analysis suggests that moderate increases in the sensitivity ratio of future telescopes can potentially probe the QCD axion model with $n=3.34$ and $m_a\simeq 50\,\mu$eV as well as similar axions with $n=3.34$ up to the cosmological mass band $10^{-6}\,$eV$\leq m_a \leq 10^{-4}\,$eV.
We emphasize that the detection of Bosenova signals requires broadband axion DM searches and that the exemplary CASPEr experiment using derivative couplings is of resonant type.
Our predictions thus motivate further innovation in broadband DM experiments, both for axion-photon and axion-fermion couplings. 

Our combined study of the galactic ASMF and AS accretion indicates promising possibilities for future use of galactic axion substructures as signatures of light scalar particles.
We emphasize that our approach has been conservative in the following regards:

1.~We have modelled the present-day ASMF using the linear growth MCMF from \cite{Fairbairn_2018, Maseizik:2024qly} and 
2.~using the initial core-halo relation \eqref{eq:CoreHalo}.
These predictions could easily be underestimating the number of bosenovae by neglecting the long-time accretion of MC axion DM onto the AS and by neglecting non-linear effects in the MC evolution.
To improve these results, further numerical studies involving the time- and $\Phi$-dependence of the AS-MC evolution as well as improved predictions for the non-linear growth of the MCMF would be required. 
Note that Refs.~\cite{Dmitriev:2023ipv, Chang:2024fol} employ a study of the late-time core-halo relation; however we emphasize that the effects of self-interactions, which imprint the maximum mass $M_{\star,\lambda}$ as an upper cutoff on the ASMF, have not been considered yet.

3. By taking Eq.~\eqref{eq:N_Novae}, we have not taken into account the possibility of multiple bosenovae occurring inside a single AS-MC system for simplicity. 
We do however find evidence for such systems since the self-similar accretion rates predicted by \cite{Dmitriev:2023ipv} scale as $\delta M_\star / \delta t\sim \tau_\mathrm{gr}^{-1} \sim \Phi^4$.
For the densest MCs in the MCMF with $\Phi> 10$, which dominate the contribution to $\mathcal{N}_\mathrm{Nova}$, multiple axion bursts within $t_\mathrm{obs}=1\,$yr are possible.

4. We have taken the self-similar accretion of \textit{isolated} AS-MC systems found in the numerical simulations of \cite{Dmitriev:2023ipv} to model AS accretion.
It is however likely that the host MC additionally acquires external dark matter through gravitational capturing from either the NFW background or other miniclusters.
We have suggested a similar model accounting for external accretion of \textit{non-isolated} AS-MCs systems in our companion paper \cite{Maseizik:2024qly} and emphasize the need for a combined consideration of the two effects.

The various extensions of our approach listed above have the potential to meaningfully refine our current predictions.
Other important extensions of our determination of the number of Bosenovae in Eq.~\eqref{eq:N_Novae} involve considerations of the long-time survivability of accreting AS-MC systems and possible AS reformation times after a core collapse.
In theory, and given repeated axion bursts over long timescales, an axion star could deplete a sufficiently large fraction of its host MC mass to eventually become stuck at a sub-critical mass $M_\star < M_{\star,\lambda}$.
This effect would diminish the number of accretion-induced Bosenovae at late times, especially for large $\Phi \sim 10^4$ and small $\mathcal{M}$.
Apart from the requirement that $M_{\star,\mathrm{acc}} \geq 0$ in Eq.~\eqref{eq:M_star_acc}, we have neglected such scenarios due to lack of knowledge on the time evolution and reformation times of ASs following a recent Bosenova.

We conclude that our simplified treatment of the present-day AS mass distribution of the Milky Way and its self-similar accretion provides strong evidence for the detectability of repeated bosenovae in our galaxy.
Future considerations of this scenario can likely improve our predictions by means of the extensions mentioned above.
Axion burst signals from galactic bosenovae hence provide a compelling laboratory for future searches of axion DM using existing and upcoming broadband experiments.

\section*{Acknowledgements}
We thank Jason Arakawa, Marianna Safronova, Volodymyr Takhistov, and Muhammad Zaheer for useful discussions.
We thank Benedikt Eggemeier and Virgile Dandoy for helpful discussions about miniclusters and tidal disruption.
The work of JE was supported by the Swedish Research Council (VR) under grants 2018-03641 and 2019-02337.
This work is funded by the Deutsche Forschungsgemeinschaft (DFG, German Research Foundation) under
Germany’s Excellence Strategy – EXC 2121 “Quantum
Universe” – 390833306.

\appendix

\section{Tidal Disruption of Miniclusters}

Fairbairn et al. \cite{Fairbairn_2018} showed that the cumulative mass fraction of miniclusters with overdensity $\Phi > \Phi_0$ may be well-fit by
\begin{align}
\mathcal{P}\left(\Phi>\Phi_0\right)=\frac{1}{\left[1+\left(\Phi_0 / a_1\right)\right]^{a 2}}, \label{eq:CDF}
\end{align}
with fitting parameters $a_1=1.023$ and $a_2=0.462$.
\begin{figure}[h]
    \centering
    \includegraphics[width=.49\textwidth]{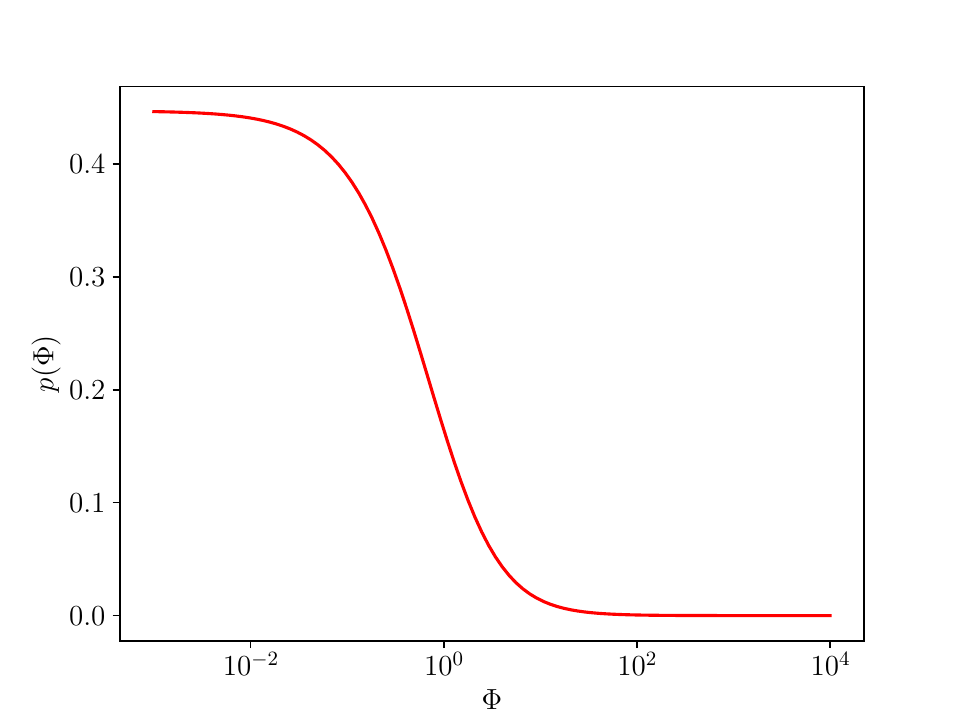}
    \caption{Probability distribution function of miniclusters with overdensity $\Phi$ obtained from the simulations in \cite{Kolb_1994} using the Pearson-VII fit \cite{Fairbairn_2018}.}
    \label{fig:CDF_Phi}
\end{figure}
This gives the probability distribution function
\begin{align}
    p_\Phi(\Phi) = \frac{a_2}{a_1 (1 + \Phi/a_1)^{a_2+1}} \label{eq:PDF}
\end{align}
of miniclusters with overdensity $\Phi$ shown in Fig.~\ref{fig:CDF_Phi}.
Note that we assume $\Phi, \mathcal{M}$ and the galactocentric coordinate $R$ to be independent parameters throughout this work.

Another effect on the MCMF that needs to be considered is given by the tidal interactions between miniclusters and astrophysical objects, mainly stars in the vicinity of the galactic center.
The resulting \textit{tidal disruption of miniclusters} in the galactic bulge constrains the galactocentric coordinate $R$ to roughly
\begin{align} \label{eq:R_surv}
    R ~ & \geq ~R_{surv} ~ \simeq 1\,\mathrm{kpc} \,,
\end{align}
which is motivated by the simulations of Ref.~\cite{Kavanagh:2020gcy, Shen_2022}.
The authors of Ref.~\cite{Kavanagh:2020gcy} found a low survival probability of $\lesssim\mathcal{O}(10)\%$ for MCs at $R < R_{surv}$ and on the other hand large survival rates $\sim90\%$ for MCs at $R > R_{surv}$.
Tidal disruption also impacts on the density parameter $\Phi$, with the survival rate
\begin{align}
    \mathcal{P}_\mathrm{surv}(\Phi) &= \frac{1}{2} \left[1 + \tanh\left(\frac{\log_{10}\Tilde{\rho}_\mathrm{mc}(\Phi) - 4.25}{2}\right) \right] \,, \nonumber\\
    \Tilde{\rho}_\mathrm{mc}(\Phi) &= \frac{\rho_\mathrm{mc}(\Phi)}{1 M_\odot \text{pc}^{-3}} \label{eq:P_surv}\,,
\end{align}
given in terms of the dimensionless MC density $\Tilde{\rho}_\mathrm{mc}$ derived from $\rho_\mathrm{mc}(\Phi)$ in Eq.~\eqref{eq:rho_mc} \cite{Dandoy_2022}.

\section{Axion Star Mass Growth Rates}

\begin{figure}[h]
\centering
\includegraphics[width=\columnwidth]{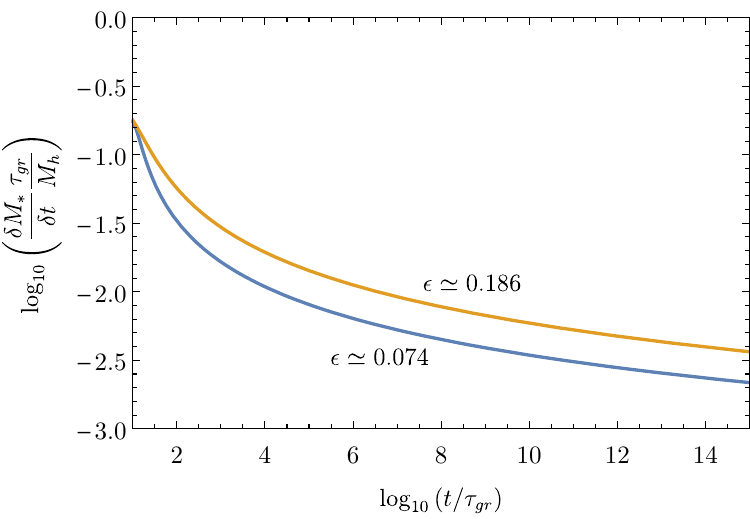}
\caption{Accretion rates for two representative simulations from \cite{Dmitriev:2023ipv} over time.}\label{fig:AccLevkov}
\end{figure}

For visualization, we have plotted the AS growth rates from Eq.~\eqref{eq:acc_Levkov} for some exemplary AS-MC systems taken from the simulations in \cite{Dmitriev:2023ipv} in Fig.~\ref{fig:AccLevkov}.
The different values of $\epsilon$ are computed from Eq.~\eqref{eq:epsilon} and thus amount to MCs with different $\mathcal{M},\Phi$ \cite{Dmitriev:2023ipv}.
Note that the accretion rates saturate for large $t \gg \tau_\mathrm{gr}$, corresponding to large relative AS masses $M_\star/\mathcal{M}$.
This observation motivates our simplified but conservative treatment of the critical AS mass $M_{\star,\mathrm{acc}}$ obtained from ongoing accretion over some time $t_\mathrm{obs}$ in Eq.~\eqref{eq:M_star_acc} as argued in the main text.

\section{Mass Functions and Cutoffs} \label{app:Cutoffs}

In order to consistently treat the different cutoffs of the MCMF following Ref.~\cite{Maseizik:2024qly}, we determine the effective low-mass cutoff for AS-MC systems reaching criticality within $t_\mathrm{obs}$ by writing
\begin{align}
    \mathcal{M}_{\lambda,\min} &\equiv \max[\mathcal{M}_{\min},\, \mathcal{M}_{h,\min},\, \mathcal{M}_{\lambda,\mathrm{acc}}(\Phi, t_\mathrm{obs})] \,,
    \label{eq:M_lambda_min}
\end{align}
given in terms of the MC masses from Eqs.~\eqref{eq:M0_cutoff}, \eqref{eq:M_h_min_CoreHalo} and \eqref{eq:M_acc}.
While typical MCs with $\Phi \sim 1$ have $\mathcal{M}_{\lambda,\min}\simeq \mathcal{M}_{\lambda,\mathrm{acc}}(\Phi, t_\mathrm{obs})$, the densest ones with $\Phi \sim 10^4$ can have sufficiently large accretion rates for the low-mass cutoffs of the MCMF to become relevant.
Note that in Eq.~\eqref{eq:M_lambda_min} we have neglected the AS radius cutoff, since for the densest MCs with $\Phi\sim 10^4$, the validity of the canonical core-halo relation \eqref{eq:CoreHalo} used to derive the radius cutoff in Ref.~\cite{Maseizik:2024qly} is arguably violated.
A direct application of the radius cutoff would diminish our signal, because many of the densest MCs with $\Phi \gg 10$ would be removed from the sample, however the strong $\Phi$-dependence in our study implies significant uncertainty on the applicability of the radius cutoff. 
In order to resolve this uncertainty, future studies should extend the numerical analysis of the core-halo mass relation performed in Ref.~\cite{Schive:2014dra} to account for variable MC densities $\Phi$.

With the effective low-mass cutoff from Eq.~\eqref{eq:M_lambda_min}, the corresponding high-mass cutoff used in Eq.~\eqref{eq:N_Novae} can be set to 
\begin{align} \label{eq:M_lambda_max}
    \mathcal{M}_{\lambda,\max}=\min(\mathcal{M}_{\max},\mathcal{M}_\lambda) \,,
\end{align}
considering the case where $\mathcal{M}_{\max}<\mathcal{M}_\lambda$ does not reach the critical value initially.

\bibliographystyle{apsrev4-2}
\bibliography{references.bib}

\end{document}